\documentclass[a4paper, 12pt]{article}

\usepackage[top=1in, bottom=1in, left=1in, right=1in]{geometry}
\usepackage{setspace}
\usepackage{natbib}
\bibpunct[, ]{(}{)}{;}{a}{,}{,}

\usepackage{pgfplots}
\usepackage[]{endfloat}

\usepackage{hyperref}
\usepackage{multirow}
\usepackage{booktabs}

\usepackage[colorinlistoftodos]{todonotes}

\usepackage{amsmath}

\title{Bayesian inference for multistate `step and turn' animal movement in continuous time}
\author{Alison Parton and Paul G. Blackwell}
\date{2017}

\begin{document}
\let\B\boldsymbol

\maketitle

\begin{abstract}
Mechanistic modelling of animal movement is often formulated in discrete time despite problems with scale invariance, such as handling irregularly timed observations. A natural solution is to formulate in continuous time, yet uptake of this has been slow. This lack of implementation is often excused by a difficulty in interpretation. Here we aim to bolster usage by developing a continuous-time model with interpretable parameters, similar to those of popular discrete-time models that use turning angles and step lengths. Movement is defined by a joint bearing and speed process, with parameters dependent on a continuous-time behavioural switching process, creating a flexible class of movement models.

Methodology is presented for Markov chain Monte Carlo inference given irregular observations, involving augmenting observed locations with a reconstruction of the underlying movement process. This is applied to well known GPS data from elk (\emph{Cervus elaphus}), which have previously been modelled in discrete time. We demonstrate the interpretable nature of the continuous-time model, finding clear differences in behaviour over time and insights into short term behaviour that could not have been obtained in discrete time.
\end{abstract}
     
\section{Introduction}
\label{Section:Intro}
The study of individual animal movement is an active area of ecological research, with advances in tracking technologies allowing data collection at increasing precision and frequency. This ability to capture short term movement has motivated the study of different movement behaviours presented by an animal over time. A number of statistical methodologies have been applied to attempt to tackle questions such as: the number of behavioural modes present, when/how often transitions between these occur, and the characteristics of movement they represent. Recent applications include, for example: \citet{Kuhn2009,McEvoy2015,McKellar2015}. 

Modelling approaches can be classified by their formulation of time; continuous models define movement at any positive, real time, whereas discrete models are defined only on some pre-determined `grid' of times. Often, the time scale in a discrete analysis is that given by the sampling scheme of the observations, leading to problems regarding irregular or missing observations~\citep{Patterson2016}, along with concerns regarding suitability and interpretability~\citep{Codling2005,Rowcliffe2012,Nams2013,Harris2013}. This lack of scale invariance places unwarranted importance on the chosen time frame, suggesting no way to combine multiple sources of data or compare analyses. Further, if a discrete-time model is thought of as observations from a continuous-time process, the existence of such a process and the effect of discretisation is not trivial to address. For example, not all discrete-time Markov chains have a continuous-time counterpart. Continuous-time models can therefore be seen as the `gold-standard' of movement modelling, avoiding these challenges through being scale invariant and respecting the continuous nature of an animal's movement.  

The continuous-time model of \citet{Johnson2008} adopts the popular movement assumption of a correlated random walk, modelling velocity via a stochastic differential equation and using a state space framework to incorporate observation error. The ability to incorporate behavioural switching however, is limited, either being highly restricted (setting velocity to zero for a stationary state at known times based on additional tag information~\citep{Johnson2008}), or simplifying to a discrete-time behavioural process~\citep{Hanks2011,McClintock2014} or movement process~\citep{Breed2012}. Similarly, the correlated and biased movement models of \citet{Kranstauber2014} use discrete-time methods for estimating the behavioural process. \citet{Blackwell2015} overcome these limitations by modelling location and allowing for a rich class of behavioural processes dependent on both environmental covariates and time via continuous-time Markov chains. A set of models able to incorporate a range of movement assumptions including the home range movement of \citet{Blackwell2015} are given in \citet{Fleming2014}, basing inference on the semivariance function of the underlying movement. This approach offers a flexible range of models, but the user is unable to associate behaviours directly with environmental information or identify the behavioural state of the animal at a specific point in time. The functional model of \citet{Buderman2016} fits splines to infer movement in continuous time, offering much versatility. However, as the estimable quantities of this approach are parameters of splines, rather than mechanistic parameters such as a `mean speed', the interpretation of these quantities is unclear. A recent generalisation using basis functions by \citet{Hooten2016} is a promising development, able to incorporate a wide range of movement and observation error. An alternative approach to those above is given by~\citet{Hanks2015} in which movement is defined in discrete space, using a Markov chain to model location switches. The inference method they propose, however, requires imputing continuous-time movement paths via some other movement model (examples include \citealp{Johnson2008} and \citealp{Buderman2016}), therefore inheriting such a model's associated assumptions and limitations.

The uptake of continuous-time approaches has been somewhat limited, owing in part to the difficulty for the practitioner to interpret the estimated instantaneous movement and behavioural parameters~\citep{McClintock2014}. In contrast, a class of discrete-time movement models based on `step lengths' and `turning angles'~\citep{Kareiva1983,Morales2004} attract widespread use~\citep{McClintock2012}. The behaviour of the animal is assumed to follow a Markov chain, with movement evolving according to behaviour-specific parameters. Within a behaviour, movement is defined by the straight line `step length' between two consecutive locations and the `turning angle' between three consecutive locations, following parametric distributions such as the Weibull and the wrapped Cauchy, respectively~\citep{Morales2004,McClintock2014}. Popular variants on this include state space models to incorporate observation error~\citep{Patterson2010, Jonsen2013}, hidden Markov models for efficiency~\citep{Langrock2012} and change point analysis rather than Markov chains to identify behavioural switches~\citep{Gurarie2009,Nams2014}.

\citet{Parton2016} introduce a continuous-time movement model based on similar quantities to those of the popular discrete-time `step and turn' models. This provides familiar descriptive parameters for estimation, whilst respecting the inherent continuous-time characteristic of movement, having the ability to handle missing and irregular observations with ease. The inference method involves simulating realisations of the underlying movement trajectory at a finer time scale than that observed, furthering our goal of providing easily understood movement analysis through the ability to visualise and relate estimated parameters to the movement they describe. This method is demonstrated on noisy observations of a reindeer (\emph{Rangifer tarandus}), taken at mostly 2~minute intervals. In Figure~2 of \citet{Parton2016}, the examples of reconstructed movement paths highlight that the characteristics of movement inferred from the observations are markedly different from a simple linear interpolation of such observations. Without accounting for observation error, as in many discrete-time methods, linearly interpolating between observations would lead to a small number of large ($\pm \pi$) turning angles. To account for these, inference would describe movement that is tortuous (correlated random walk with low correlation). However, if observation error is accounted for, \citet{Parton2016} show that the information provided by all the observations suggests movement that is persistent (correlated random walk with high correlation).

Describing only single state movement limits \citet{Parton2016} to applications with short term sampling periods. Our aim here is to introduce a statistical, multistate movement model in continuous time able to provide intuitive and easily interpretable estimated parameters for the non-statistical user. Multistate switching movement is introduced by extending \citet{Parton2016} to include a continuous-time Markov chain behavioural process. Section~\ref{Section:Model} introduces our proposed model and an approach for fully Bayesian inference given observed telemetry data is outlined in Section~\ref{Section:Algorithm}. The interpretability of this method is demonstrated in Section~\ref{Section:Elk} on well known GPS data from a single elk (\emph{Cervus elaphus}).

\section{Multistate movement based on steps and turns}
\label{Section:Model}

\subsection{Single state movement model}
\label{Subsection:MoveModel}
The basic component for movement follows that of \citet{Parton2016}, in which the animal has both a bearing $\theta(t)$ and a speed $\psi(t)$ at time $t \geq 0$. The bearing process describes the direction the animal is facing, assumed to evolve according to Brownian motion with volatility $\sigma_\theta^2$ so that
\[
\mathrm{d} \theta(t) = \sigma_\theta \mathrm{d} W(t),
\]
where $W(t)$ is the Wiener process~\citep{Guttorp}. This reflects the common assumption of persistence, where the animal will most likely travel in the same direction over a short period of time. Over a finite period of time, the change in direction of facing will be a wrapped Gaussian with mean zero and a variance which is a linear function of time.

The direction an animal is facing at any time is constrained to $[-\pi,\pi]$, however, here $\theta(t)$ is not constrained in this way and can take any real value. For example, given times $0\leq t < s$, let $\theta(t)=0$ and $\theta(s)=2\pi$. Although the animal was facing the same direction at both times, there is information about the behaviour of the process between these points, as the animal has turned an entire `loop' over this time frame (with the distribution of this constrained process being a Brownian bridge)

A one-dimensional Ornstein-Uhlenbeck process~\citep{Iacus} is assumed to govern the speed with which the animal is travelling, with parameters $\lbrace \mu, \beta, \sigma_\psi^2 \rbrace$ so that
\[
\mathrm{d} \psi(t) = \beta(\mu - \psi(t)) \mathrm{d} t + \sigma_\psi \mathrm{d} W(t).
\] 
Hence the animal's speed is stochastic but correlated, with long-term average $\mu$ and variance $\sigma_\psi^2/2\beta$.

Alternate modelling assumptions to those presented may be desired dependent upon application. A more direct comparison with discrete-time correlated random walk models would be to model speed as Brownian motion so that distances travelled over disjoint time periods are independent. Similarly, directed/biased movement could be achieved by altering the Brownian motion on the bearing process, or assuming some Ornstein-Uhlenbeck process.

The joint process given by the bearing and speed of the animal completely defines the location process $\B{Z}=\lbrace \B{X},\B{Y} \rbrace$, given by
\[
	\mathrm{d} X(t) 
= \psi(t)\cos(\theta(t)), 
	\quad
	\mathrm{d} Y(t) 
= \psi(t)\sin(\theta(t)).
\]

\subsection{Multistate switching model}
\label{Subsection:BehavModel}
To reflect the changing behaviours of an animal over time, a switching model is employed, with different movement characteristics for each state~\citep{Blackwell1997,Morales2004,McClintock2012,Blackwell2015}. The behavioural process is taken to be a continuous-time Markov chain with switching rates $\B{\lambda}$ and probabilities $\B{q}$~\citep{Guttorp}. The animal will follow behavioural state $i$ for a length of time exponentially distributed with rate $\lambda_i$, before switching to state $j$ with probability $q_{i,j}$. Within a behaviour there is a corresponding set of parameters describing the movement, as in Section~\ref{Subsection:MoveModel}. With this extension in place the marginal joint process of bearing and speed is not Markovian, however the joint process of behaviour, bearing and speed is. The movement of the animal is therefore parametrised by the set $\B{\Phi} = \lbrace \B{\Phi}_B, \B{\Phi}_M \rbrace$, with $\B{\Phi}_B= \lbrace \lambda_i, q_{i,j} \rbrace$ and $\B{\Phi}_M= \lbrace \sigma_{\theta,i}^2, \mu_i, \beta_i, \sigma_{\psi,i}^2 \rbrace $ for $i \neq j \in \lbrace 1,\ldots,n \rbrace$, where $n$ is the number of behavioural states. 

\subsection{Simulating multistate movement}
\label{Subsection:SimMove}
\begin{figure}
	\includegraphics[width=\linewidth]{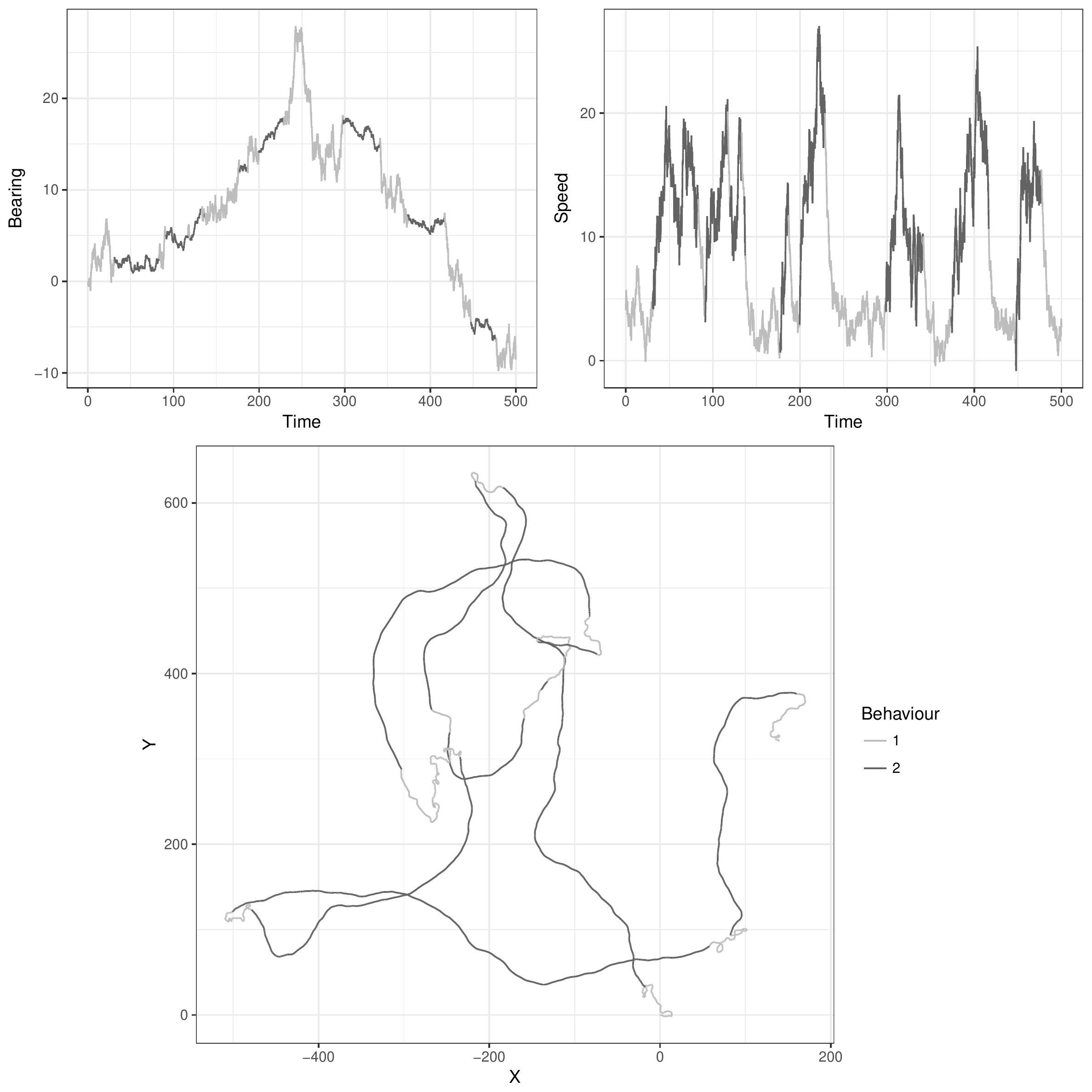}
	\caption[Example of a simulated movement path]{An example of a simulated movement path with two behavioural states. The simulated bearing and speed process are shown, coloured by the simulated behavioural process, along with the resulting 2-dimensional locations.}
	\label{Fig:SimEx}
\end{figure}

Realisations of movement given parameters $\B{\Phi}$ can be easily simulated, with an example of such in Figure~\ref{Fig:SimEx}. The behavioural process is simulated according to a continuous-time Markov chain with generator matrix defined by $\B{\Phi}_B$. Given a current behaviour $B(t)=s$, this involves drawing the time until the next behavioural switch from an exponential distribution with rate $\lambda_{s}$ and then choosing the new behaviour $j \neq s$ with probability $q_{s,j}$.

Given a realisation of the behavioural process, movement is simulated at an approximate time scale $\delta t$, which can be arbitrarily fine. If the behaviour at time $t$ is $B(t)=s$, then the bearing and speed are given as
\begin{align} \label{Eq:SimBearingSpeed}
\theta(t+\delta t) \ &| \ \theta(t), s \sim \text{N}\left( \theta(t), \ \sigma_{\theta,s}^2 \delta t \right), \\
\psi(t+\delta t) \ &| \ \psi(t), s \sim \text{N} \left( \mu_s + \exp\lbrace-\beta_s \delta t\rbrace(\psi(t)-\mu_s), \ \frac{\sigma_{\psi,s}^2}{2\beta_s} \left(1-\exp\lbrace-2\beta_s\delta t\rbrace \right) \right). \label{Eq:SimSpeed}
\end{align}
Given this approximation, the familiar notion of a `step' is recovered by $\nu(t)=\psi(t)\delta t$.

Given the joint processes $\lbrace \B{\theta},\B{\nu} \rbrace$, the Euler-Maruyama approximation of location in 2-dimensional space is given by the cumulative sums
\begin{equation} \label{Eq:LocSum}
X(t_i) = X(t_0) + \sum_{j=1}^{i-1} \nu(t_j) \cos(\theta(t_j)), \quad Y(t_i) = Y(t_0) + \sum_{j=1}^{i-1} \nu(t_j) \sin(\theta(t_j)).
\end{equation}

\section{The Markov chain Monte Carlo algorithm}
\label{Section:Algorithm}
Observations $\B{Z}$ of an animal's 2-dimensional location are taken at a finite, but irregular, series of times $\B{t}$. The likelihood of these observations given parameters $\B{\Phi}$ is intractable due to the complicated relationship between the locations and parameters when the bearing and speed processes are unobserved. This is further complicated by the unobserved behavioural process, where there is the possibility of multiple switches between observations. The following describes the Markov chain Monte Carlo algorithm used to carry out inference given observations. 

Following \citet{Blackwell2003} a data augmentation approach is taken, simplifying the relationship between observations and parameters by augmenting the data with the times of all behavioural switches. Here, augmentation also includes an approximation to the underlying bearing and speed processes on some (arbitrarily fine) time scale. The hybrid Markov chain Monte Carlo algorithm used splits the quantities of interest into three groups to update separately, in each case conditional on all other quantities. In cases where the full conditional distribution can be directly sampled from, Gibbs sampling is employed, and in all other scenarios the Metropolis-Hastings sampler is used (see e.g.~\citet{Gelman} for general sampling methods). The groups to be separately sampled from are the behavioural parameters ($\B{\Phi}_B$), the movement parameters ($\B{\Phi}_M$), and the unobserved refined path consisting of behavioural switches, bearings and speeds ($\B{B},\B{\theta},\B{\nu}$).

Sections~\ref{Subsection:BehavParam} and~\ref{Subsection:MoveParam} describe the sampling schemes used for the behavioural and movement parameters, respectively. In both cases the sampling is standard, employing Gibbs sampling and a random walk Metropolis-Hastings algorithm. Section~\ref{Subsection:Path} describes the Metropolis-Hastings algorithm used for the reconstruction of the unobserved refined path, in which a novel method of simulation is used to create the independent proposals within this sampling scheme.
 
\subsection{Sampling the behavioural process parameters}
\label{Subsection:BehavParam}
The behavioural process parameters are sampled conditional on the complete observation of the behavioural process. Conjugate distributions for the switching rates ($\B{\lambda}$) and probabilities ($\B{q}$) of a continuous-time Markov chain are gamma and Dirichlet, respectively. Assuming such conjugate priors allows direct sampling from the posterior conditional as a Gibbs steps~\citep{Blackwell2003}. Further details are given in Appendix~\ref{AppSubsec:BehavParam}.

\subsection{Sampling the movement process parameters}
\label{Subsection:MoveParam}
The movement process parameters are sampled conditional on the complete observation of the refined path (both behaviour and movement) and the behavioural parameters. The movement parameters are updated simultaneously using a random walk Metropolis-Hastings step, with independent proposals for each parameter. Since all movement parameters are constrained to be positive, independent univariate Gaussians truncated below at zero are used as proposal distributions to generate the step in the random walk.

In a simultaneous update of the movement parameters, the likelihood of the refined movement path is calculated for the current and proposed parameters and combined with the appropriate prior probability. The standard Metropolis-Hastings acceptance ratio is used to decide on the acceptance of the proposal. Further details are given in Appendix~\ref{AppSubsec:MoveParam}.

\subsection{Reconstructing the unobserved refined path}
\label{Subsection:Path}
\begin{figure}
	\includegraphics[width=\linewidth]{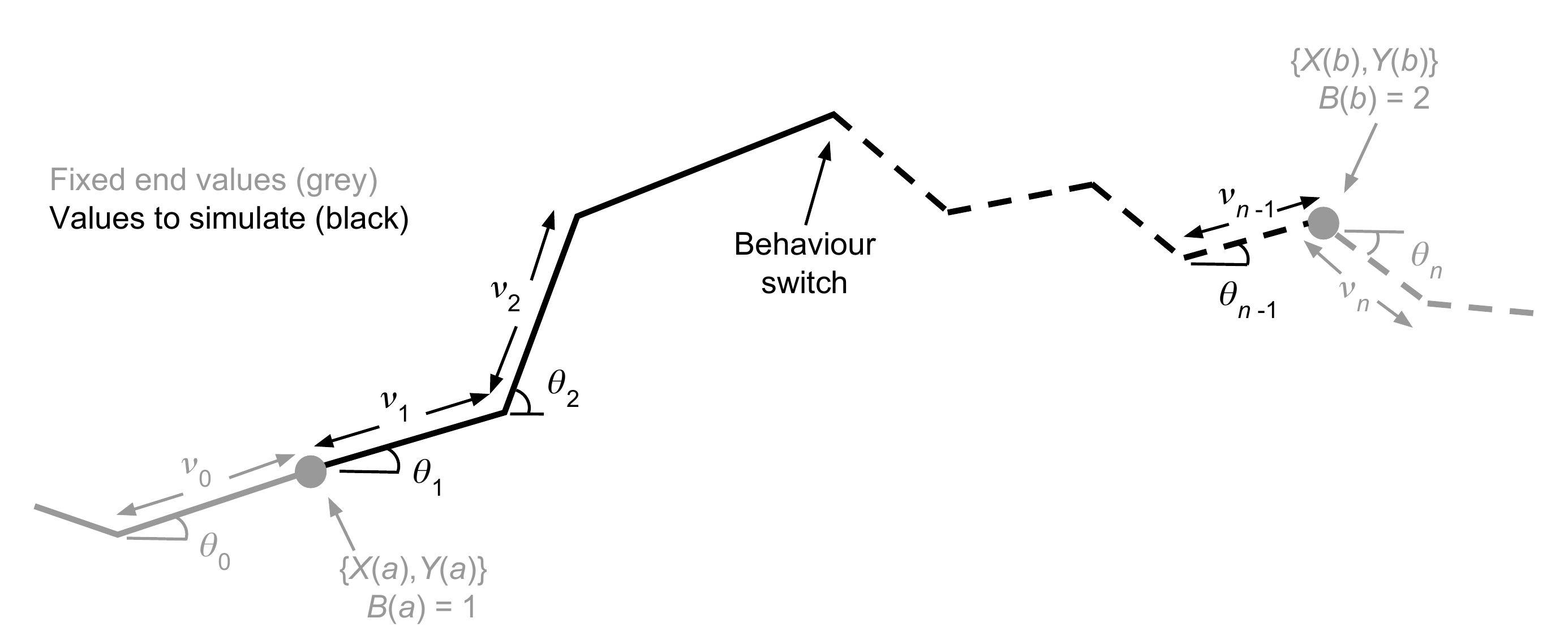}
	\caption[Diagram of refined path section]{Diagram of a section of the refined path, with fixed endpoint locations at the times $a$ and $b$. The behavioural process, $B$ (represented as two states with solid and dashed lines here), is simulated with fixed endpoints $\lbrace B(a),B(b)\rbrace$. The bearing and step processes, $\lbrace \theta_1,\ldots,\theta_{n-1}, \nu_1,\ldots,\nu_{n-1} \rbrace$, are simulated given fixed endpoints $\lbrace \theta_0,\theta_n,\nu_0,\nu_n\rbrace$.}
	\label{Fig:PathDiagram}
\end{figure}

The key step for inference is to sample the unobserved `refined path'---given by the behavioural process, and the bearing and speed processes at a refined time scale---conditional on the parameters. As the dimension of the full movement path will be large (the example of Section~\ref{Section:Elk} leads to a path with around 2,300 locations at the chosen refined time scale), reconstruction is carried out on random short sections. The aim is to simulate the refined path between two observation times $a$ and $b$, conditional on the fixed path outside of these times and a set of parameters. This can easily be extended to span multiple observed locations. A diagram of this scenario is given in Figure~\ref{Fig:PathDiagram}, with two circular points showing the fixed observations that the path will be simulated between. 

The quantities to simulate are those in black in Figure~\ref{Fig:PathDiagram} consisting of: the behavioural process $\B{B}$ between times $a$ and $b$, the bearings $\lbrace \theta_1,\ldots,\theta_{n-1}\rbrace$ and the steps $\lbrace\nu_1,\ldots,\nu_{n-1}\rbrace$. The fixed values that are to be conditioned upon are displayed in grey in Figure~\ref{Fig:PathDiagram} consisting of: the locations $\lbrace\B{Z}(a), \B{Z}(b)\rbrace$, the behaviours $\lbrace B(a),B(b)\rbrace$, the bearings $\lbrace\theta_0,\theta_n\rbrace$ and the steps $\lbrace\nu_0,\nu_n\rbrace$. As the bearing and step processes are given by a discrete-time approximation, the fixed points are the values of the respective process at the refined point immediately before and after the path section of interest, as in Figure~\ref{Fig:PathDiagram}. 

Simulating the quantities of interest conditional on all fixed values is not possible due to the non-linearity of the location process (see Equation~\ref{Eq:LocSum}), and so a proposal path section is simulated from a simpler distribution that is then accepted or rejected using a Metropolis-Hastings ratio. An independence sampler is employed using a novel simulation method to propose a new path section, described below. Further details on the acceptance condition is given in Appendix~\ref{AppSubsec:Path}.

\subsubsection{Simulating a refined path proposal}
\label{Subsection:PathPropMethod}
A behavioural proposal $\B{B}^*$ is simulated between the times $a$ and $b$, given fixed values $\lbrace B(a), B(b)\rbrace$ and parameters $\B{\Phi}_B$, by a rejection method. A continuous-time Markov chain with parameters $\B{\Phi}_B$ starting at $B(a)$ at time $a$ and ending at time $b$ is simulated (see Section~\ref{Subsection:SimMove}). If the final state is not equal to $B(b)$, then the proposal is instantly rejected. Otherwise, the path proposal continues (still with the possibility of rejection in the Metropolis-Hastings step). Less naive approaches to this simulation could be implemented (see e.g.~\citet{Hobolth2009,Rao2013,Whitaker2016}), however this naive method performed well in our examples.

Given the behavioural simulation, the set of refined times $\lbrace t_1=a,\ldots,t_{n-1}\rbrace $ is created. This must be a sequence of times between $a$ and $b$ that includes behavioural switch times, and is chosen to approximately be on some time scale $\delta t$, the choice of which is discussed in Section~\ref{Section:Discussion}. This forms the times to simulate the bearings and speed over, as in Figure~\ref{Fig:PathDiagram}.

The bearing proposal $\B{\theta}^*$ over the times $\lbrace t_1, \ldots, t_{n-1}\rbrace$ is simulated conditional on the fixed bearings $\lbrace \theta_0,\theta_n\rbrace$ at the times $\lbrace t_0,t_n=b\rbrace$, the behaviours $\B{B}^*$ and the parameters $\B{\Phi}$. The distribution of this process is a Brownian bridge with time-varying volatility parameter, dependent on behaviour. The times $\lbrace t_1, \ldots, t_{n-1}, t_n \rbrace$ are transformed, weighted by the turn volatility at each respective time, to give a process with constant volatility. The Brownian bridge is then simulated on the transformed times $\lbrace t_1^{'}, \ldots, t_{n-1}^{'}\rbrace$, given the values $\lbrace \theta_0, \theta_n \rbrace$ at the end times $\lbrace t_0,t_n^{'} \rbrace$ (see \citet{Iacus} for Brownian bridge simulation).

\paragraph{Simulating the step proposal}
To propose the steps $\B{\nu}^*$ over the times $\lbrace t_1, \ldots, t_{n-1}\rbrace$, the joint distribution of $\B{\nu}$ and $\B{Z}(b)$, given by 
\begin{equation} \label{Eq:JointStepLoc} 
\begin{pmatrix}
\B{\nu} \\ \B{Z}(b)
\end{pmatrix}
\ | \ \B{\Phi}, \B{B}^*, \B{\theta}^*, \mathcal{F} \sim \text{N}\left(
\begin{pmatrix}
\B{m}_1 \\ \B{m}_2
\end{pmatrix},
\begin{pmatrix}
\Sigma_1 & \Sigma_{1,2} \\ \Sigma_{1,2}^\text{T} & \Sigma_2 
\end{pmatrix}
\right),
\end{equation}
where $\mathcal{F} = \lbrace \B{Z}(a), B(a), B(b), \theta_0, \theta_n, \nu_0, \nu_n \rbrace$, is first constructed. The marginal distribution of $\B{\nu}$ (dimension~$n-1$) given a known behavioural process and fixed end steps is $\text{N}\left(\B{m}_1,\Sigma_1\right)$ (discussed further below). The location $\B{Z}(b)$ is given by $\B{Z}(a)+A\B{\nu}$, where
\[
A = \begin{pmatrix}
\cos(\theta_1^*) & \cdots & \cos(\theta_{n-1}^*) \\
\sin(\theta_1^*) & \cdots & \sin(\theta_{n-1}^*)
\end{pmatrix}.
\]
The marginal distribution of $\B{Z}(b)$ (dimension 2) is $\text{N}\left(\B{m}_2, \Sigma_{2}\right)$, and $\Sigma_{1,2}$ is the $(n-1)\times 2$ covariance between the steps $\B{\nu}$ and the location $\B{Z}(b)$. Given $\B{m_1},\Sigma_1,A$, values for $\B{m}_2, \Sigma_2, \Sigma_{1,2}$ can be easily calculated due to $\B{Z}(b)$ being a linear combination of the normally distributed $\B{\nu}$.

The form of $\B{m}_1,\Sigma_1$ arise from the speed process (from which $\B{\nu}$ is derived) being an Ornstein-Uhlenbeck bridge with inhomogeneous parameters, calculated by the following method. The fixed values $\nu_0, \nu_n$ are transformed to give speeds $\psi_0 = \nu_0/\delta t_0$ and $\psi_n=\nu_n/\delta t_n$.
The joint distribution $\psi_1,\ldots,\psi_n \ | \ \psi_0, \B{B}^*$ is created by iteratively applying
\begin{equation}
\psi_i \ | \ \psi_{i-1}, B(t_i) \sim \text{N}(\mu,\sigma^2),
\end{equation} 
where $\mu,\sigma^2$ are given by Equation~\ref{Eq:SimSpeed}. This joint distribution is then partitioned into $\psi_1,\ldots,\psi_{n-1}$ and $\psi_n$ in order to condition upon the known value for $\psi_n$ using standard conditioning of a multivariate normal~\citep{Eaton} to give the joint distribution $\psi_1,\ldots,\psi_{n-1} \ | \ \psi_0,\psi_n,\B{B}^*$. This distribution can be transformed back to steps $\nu_1,\ldots,\nu_{n-1}$ to give $\B{m}_1,\Sigma_1$ through a transformation by multiplying the speeds $\psi_1\ldots,\psi_{n-1}$ by the times $\delta t_1,\ldots,\delta t_{n-1}$.  

The step proposal $\B{\nu}^*$ is simulated by further conditioning $\B{\nu}$ in Equation~\ref{Eq:JointStepLoc} on the known $\B{Z}(b)$ by standard conditioning of a normal distribution~\citep{Eaton}, given by
\[
\B{\nu} \ | \ \B{\Phi}, \B{B}^*, \B{\theta}^*, \mathcal{F}, \B{Z}(b) \sim \text{N}
\left( 
\B{m}_1 + \Sigma_{1,2} \Sigma_2^{-1} \left(\B{Z}(b) - \B{m}_2\right),
\Sigma_1 - \Sigma_{1,2} \Sigma_2^{-1} \Sigma_{1,2}^\text{T}
\right).
\]
The steps are being conditioned upon a linear constraint (the fixed $\B{Z}(b)$), leading to a singular distribution. Simulation of such follows the `conditioning by Kriging' procedure in \citet{Rue}, by first simulating from the unconditioned $\B{x}\sim\text{N}(\B{m}_1,\Sigma_1)$ and adjusting for the constraint by
\[
\B{\nu}^* = \B{x} - \Sigma_{1,2} \Sigma_2^{-1}(A\B{x}-\B{Z}(b)).
\]

This path proposal method does not take into account the fixed location at the end of the section when simulating the behaviours and bearings. Therefore, a Metropolis-Hastings step (ratio details in Appendix~\ref{AppSubsec:Path}) assesses whether this proposal is accepted.

\section{Two state switching movement in elk}
\label{Section:Elk}
\begin{figure}
\includegraphics[width=\linewidth]{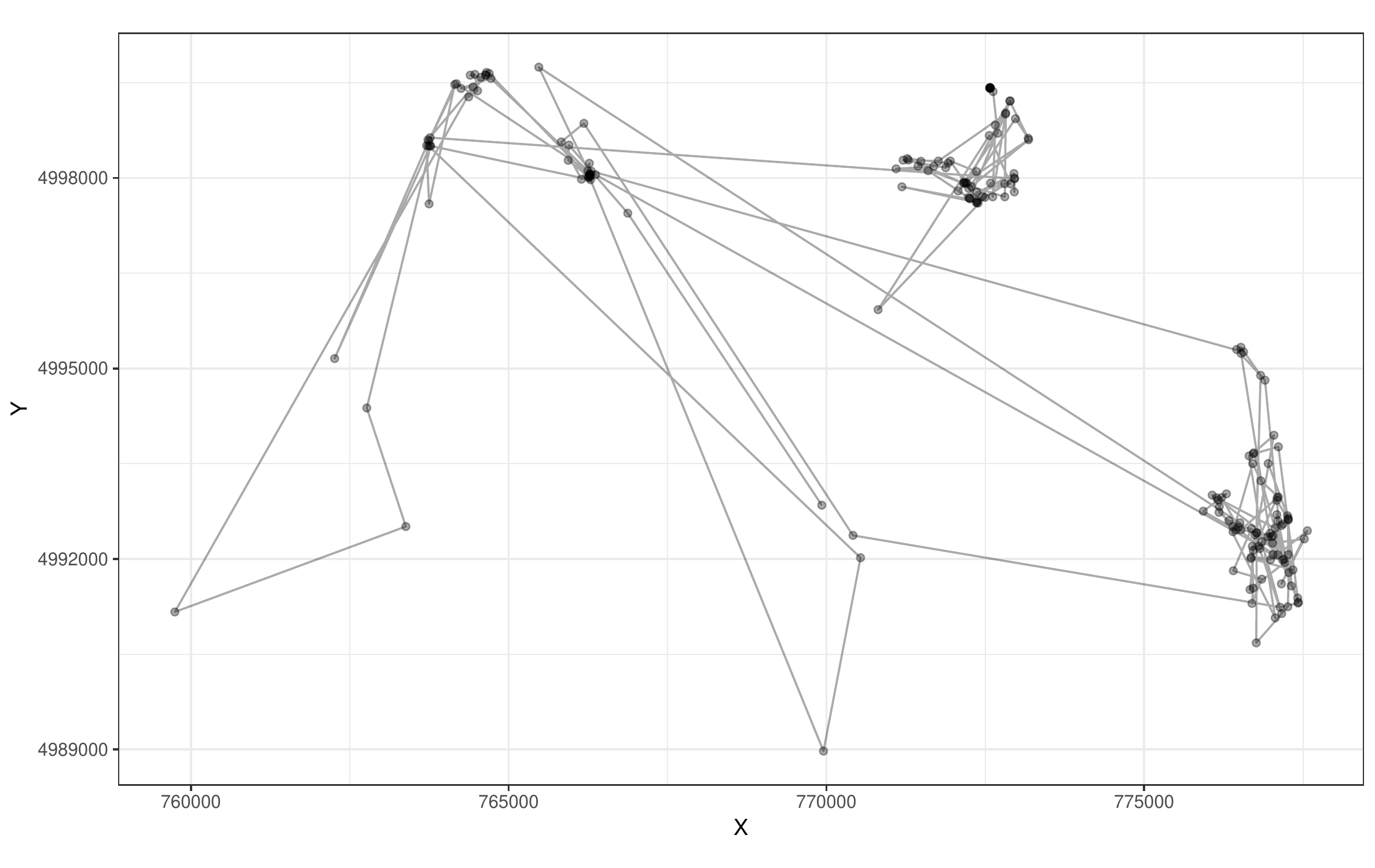}
\caption[Observed locations of elk-115]{Observed daily observations of Elk-115 (points linked chronologically with lines). Note that observed points are displayed here with transparency to highlight the times where multiple observations were captured in the same/similar location.}
	\label{Fig:ElkObs}
\end{figure}

A set of 194 daily GPS observations from the elk (\emph{Cervus elaphus}) tagged as `elk-115' are used in this example (see \url{https://bitbucket.org/a_parton/elk_example}). These observations were introduced and modelled as part of a larger set consisting of four elk in the discrete-time `step and turn' model of \citet{Morales2004}, and more recently modelled in the vignette of the \texttt{R}~package \texttt{moveHMM}~\citep{MoveHMM} applying the hidden Markov model of \citet{Langrock2012}. Observations are shown in Figure~\ref{Fig:ElkObs}, appearing to display two distinct movement modes: slow, volatile movement where observations are over-plotted, and fast, directed movement. 

\citet{Morales2004} fit a number of models to the larger dataset containing the observations from elk-115, with the model most similar to ours being the `double switch' model. Fixed switching probabilities between the two states were modelled, governing a mixture of correlated random walks. In the vignette of \texttt{moveHMM} the larger dataset is used to demonstrate a two state hidden Markov model with switching dependent on environment. For comparison with the methods here, the reproduction of analysis shown in Figure~\ref{Fig:ElkBehav} does not include this environmental information and so is the same underlying movement model as the `double switch' in \citet{Morales2004}. In both these discrete-time applications, `travelling' and `foraging' states were identified as having mean daily turning angles of close to zero and $\pi$, respectively. The implications of turn distributions not centred at zero are discussed in Section~\ref{Section:Discussion}. 

In this example, the model of Section~\ref{Section:Model} with two behaviours is applied to the elk-115 observations. The original analysis in \citet{Morales2004} described observations as being mostly daily, but with some taken at 22 and 26~hour intervals. In order to handle this irregularity, they divided the observed straight line step lengths by the sampling time frame to approximate daily steps. A method transforming the observed turning angles to some daily approximation is unclear, and so these remained as the observed values in their analysis. The open-access version of the elk data does not include the times of the observations, and rounding of the \citet{Morales2004} `daily step lengths' meant that the original observation times could not be ascertained. The analysis carried out here therefore followed that in the vignette of \texttt{moveHMM}, using the observed locations, but assuming that these were all at 24~hour intervals. The continuous-time formulation of our model, however, would easily allow for these irregularly timed observations (and missing observations, if applicable) to be handled if exact observation times were known.

Applying our presented methodology to multiple animals in the same way as \texttt{moveHMM}, by pooling information across individuals and estimating a set of population parameters, could be implemented by a simple extension to the current \texttt{R} code but is not attempted here for simplicity. Following \citet{Morales2004} and the vignette of \texttt{moveHMM}, observation error is assumed to be negligible here (though see Section~\ref{Section:Discussion}). Interest thus involves inference on the eight movement parameters, consisting of a bearing volatility and three speed parameters for each state. Using daily observations leaves large portions of the elk's movement unobserved, and so it is expected that the reconstructed movement paths, and thus parameters, for this example will be very uncertain. Rather than a full ecological analysis, this example is therefore included as a proof of concept for the presented methods and to highlight some of the possible dangers when analysing daily observations in discrete time. Readers are directed to \citet{Parton2016} for an example of single state movement on a dataset with a sampling scheme of 2~minutes to compare the uncertainty of movement reconstructions. 

\subsection{Prior and initial information}
\label{Subsection:PriorInitial}
A prior distribution specifying an upper bound on the ratio of the speed parameters to avoid the presence of negative speeds in both states was applied. To define state~2 as `travelling', a Gaussian prior with mean 0.05 and standard deviation of 0.1 was placed on the turn volatility. All remaining movement parameters had flat priors. The same prior was on both switching rates, being a gamma distribution with rate~4 and shape~0.1. This was chosen to limit the rate of behavioural switching, strongly discouraging switching occurring at a shorter time frame than 4~hours, with 90\% prior credible interval for residency time of ($6,7\times 10^{13}$)~hours. 
This prior is fairly vague when comparing with the posterior credible intervals (see below).  

An initial movement path was created at a time scale of 2~hours by taking an interpolating cubic spline between observations. The choice of a 2~hour time scale gives around 11~unknown locations for reconstruction between each pair of observations, thought to provide an acceptable trade-off between computational cost and approximation to continuous time (see Section~\ref{Section:Discussion} for further discussion of $\delta t$). The corresponding initial behavioural configuration was set by identifying any points on this path with speed above 100~m/h. Initial parameters were set as estimates from this initial path configuration. 

The algorithm of Section~\ref{Section:Algorithm} was applied for $48\times 10^5$~iterations, with each iteration consisting of a single parameter update and $100$~refined path updates on random sections of path with lengths ranging 4--24~points (i.e. 8--48~hours). Samples were thinned by a factor of $1000$ and the first quarter were treated as a `burn-in' period, leaving $3600$ stored samples of parameters and reconstructed refined paths. Long sub-path lengths are desirable as the proportion of path being updated is high. However, this incurs computational cost and has low acceptance due to high dimensionality. A mixture of short sub-path lengths (easily accepted) helps with mixing, following on from such a discussion in \citet{Blackwell2015}. The choice here was based on acceptance rates in pilot runs: lengths higher than 24 had too low acceptance to be feasible, and lengths of 4 allowed these short section updates that helped with mixing. 

\subsection{Results}
\label{Subsection:Results}
\begin{figure}
\includegraphics[width=\linewidth]{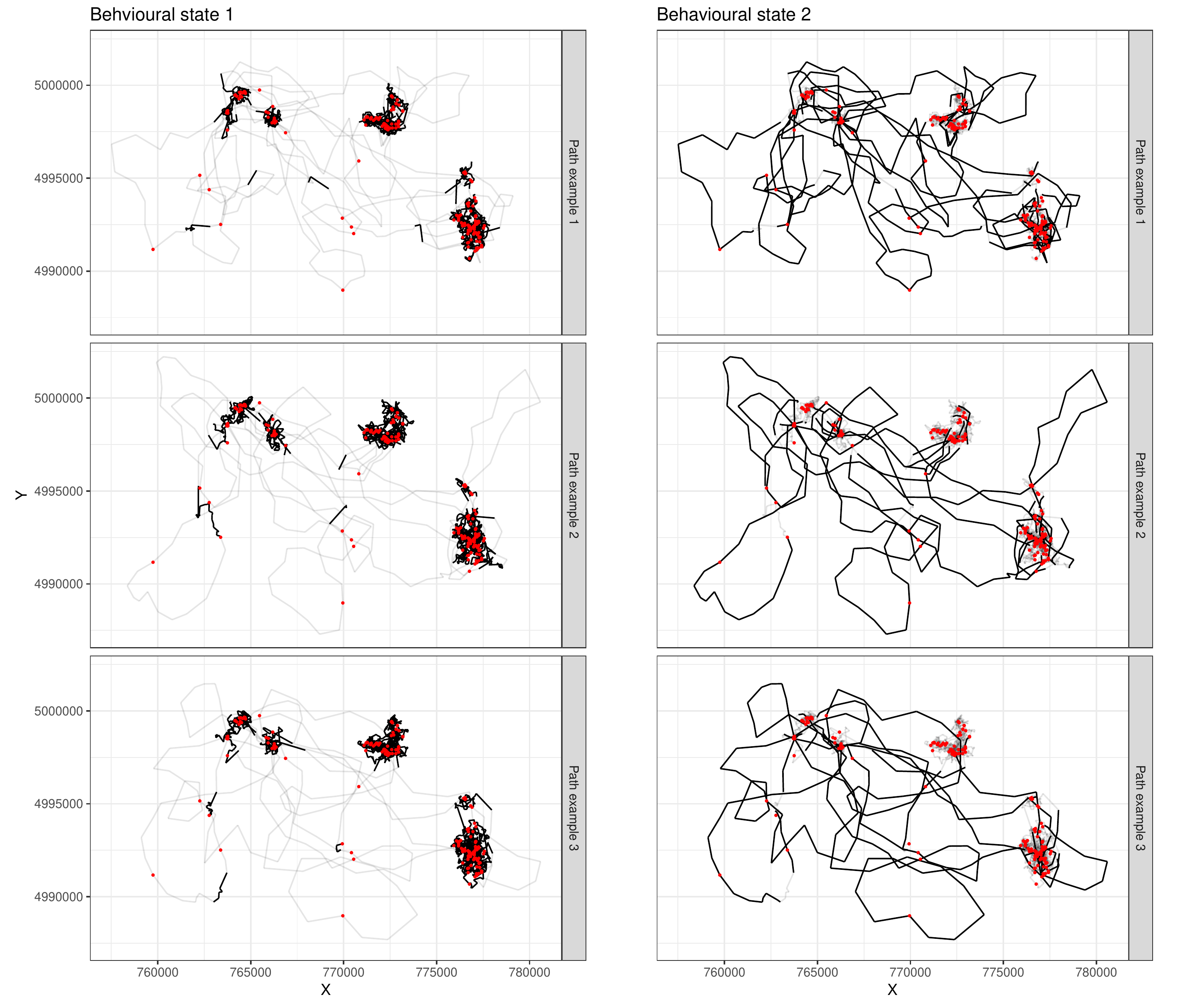}
\caption[Reconstructed path examples for elk-115]{Three examples of reconstructed refined movement paths for elk-115. For each example, the observed locations are shown as red points and the reconstructed refined path is displayed as linearly interpolated lines. The left and right panels both show the full reconstructed refined path (in grey and black), but differ by the behavioural state highlighted: the left panel highlights in black the parts of the path labelled as behavioural state~1 and the right panel highlights in black the parts of the path labelled as state~2. This separation of behavioural segments clearly highlights the difference in movement characteristics resulting from the parameters associated with the two behavioural states.}
\label{Fig:ElkPaths}
\end{figure}

Figure~\ref{Fig:ElkPaths} shows three examples (separated vertically) of the reconstructed refined movement path. Red points show the observations and the combination of grey and black lines show the three example path reconstructions. Each reconstruction is shown in two panels: the left panel highlights in black the segments of the refined path categorised as behavioural state~1, and the right panel highlights in black the segments of the path labelled as state~2. This highlights the difference in movement types between the two identified states, appearing in many ways similar in interpretation to those of \citet{Morales2004} and the vignette of \texttt{moveHMM}, having a slow `foraging' state and fast `travelling' state. These reconstructions aid in the interpretation of the movement parameters and give insight into the space use of the animal between observation times.

\begin{figure}
\includegraphics[width=\linewidth]{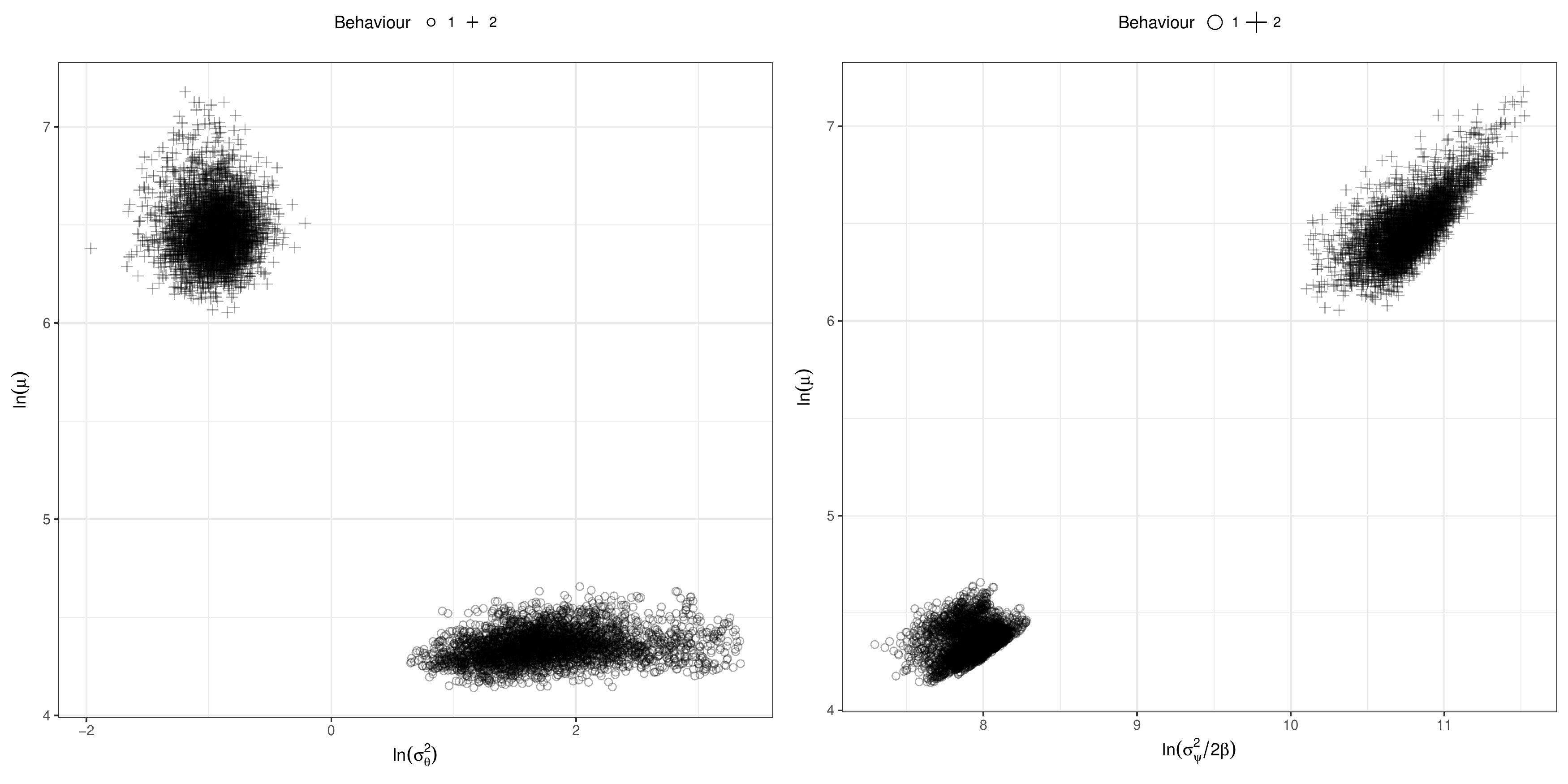}
\caption[Sample movement parameters for elk-115]{Sampled state-dependent movement parameters (on log scale) for the example using observations of elk-115. Left plot: the joint sample space between the turn volatility ($\sigma_\theta^2$) and the mean speed ($\mu$). Right plot: the joint sample space between the mean speed and the long-term speed variance ($\sigma_\psi^2/2\beta$).}
\label{Fig:ElkMove}
\end{figure}

Samples from the posterior distributions for the movement parameters, split by state, are shown in Figure~\ref{Fig:ElkMove}, showing the clear differences between the two states. Posterior summary statistics of the parameters are given in Table~\ref{Tab:ParamStat}. Behavioural state~1 has high $\sigma_\theta^2$ and low $\mu$, defining volatile, slow movement categorised here as `foraging'. The level of $\sigma_\theta^2$ for state~1 (median given by 5.61~rad per hour) is high enough to produce turns that are uniform over the sampling scheme of the observations. The median for long term travelling speed for state~1 is given by 77.3 (metres per hour). State~1 has a higher $\beta$ and lower $\sigma_\psi^2$ than state~2, describing speeds that are less correlated in the short term (the mean expression of the speed process in Equation~\ref{Eq:SimSpeed} is dominated by the first term involving the `mean speed' parameter rather than the second term involving the `current speed') and have lower variation in the long term. The movement parameters for state~1 have a low effective sample size and do not pass standard convergence diagnostics. This is due to the turn volatility being so high as produce uniform turns, and so this parameter is `drifting'.

Behavioural state~2, the `travelling' state, has low $\sigma_\theta^2$ and high $\mu$, reflecting fast, straight movement. The median long term travelling speed for state~2 is 638 (metres per hour), with speeds that are highly correlated in the short term (through a low $\beta$) but with high variation in the long term (through a high $\sigma_\psi^2$). The movement parameters for state~2 pass standard convergence diagnostics (Heidelberger and Welch) with effective sample size of over 75.

\begin{table}
\centering
\begin{tabular}{@{}lllll@{}}
\toprule
& Parameter & $5\%$ & $50\%$ & $95\%$ \\ \midrule
\multirow{5}{*}{\begin{tabular}[c]{@{}l@{}}Behaviour 1\\ (`foraging')\end{tabular}} & $\lambda_1$ (switching rate)& 0.00391 & 0.00651 & 0.0105 \\
& 
$\sigma_\theta^2$ (turn volatility) & 2.87 & 5.61 & 16.4 \\  
& 
$\mu$ (long term speed mean) & 68.8 & 77.3 & 90.2 \\ 
& 
$\beta$ & 0.627 & 1.45 & 1.94 \\ 
& 
$\sigma_\psi^2$ & 2,900 & 7,920 & 11,300 \\ 
& 
$\sigma_\psi^2 / 2\beta$ (long term speed variance) & 2,160 & 2,820 & 3,390 \\ \midrule
\multirow{5}{*}{\begin{tabular}[c]{@{}l@{}}Behaviour 2\\ (`travelling')\end{tabular}} & $\lambda_2$ (switching rate) & 0.0275  & 0.0520  & 0.0959  \\
& 
$\sigma_\theta^2$ (turn volatility) & 0.274 & 0.389 & 0.521 \\ 
& 
$\mu$ (long term speed mean) & 519 & 638 & 855 \\ 
& 
$\beta$ & 0.170 & 0.245 & 0.340 \\ 
& 
$\sigma_\psi^2$ & 16,000 & 23,600 & 29,700 \\ 
& 
$\sigma_\psi^2 / 2\beta$ (long term speed variance) & 34,300 & 47,600 & 66,400 \\ \bottomrule 
\end{tabular}
\caption[Posterior summary statistics of parameters for elk-115]{Posterior summary statistics ($5\%,50\%,95\%$ quantiles) for the sampled movement and behavioural parameters, split by state, in the elk-115 example.}
\label{Tab:ParamStat}
\end{table}

\begin{figure}
\includegraphics[width=\linewidth]{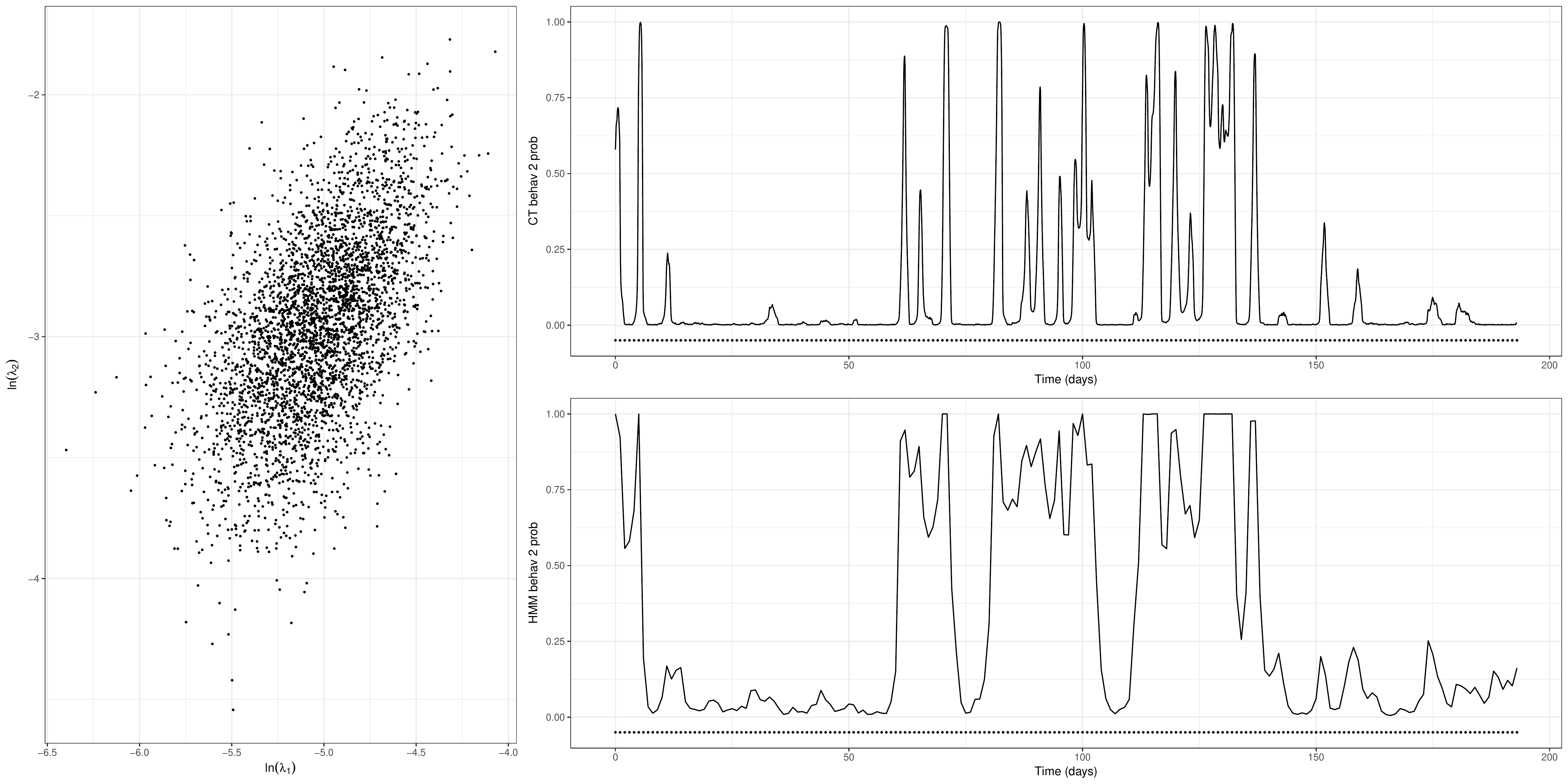}
\caption[Sampled behavioural parameters and behavioural state probabilities for elk-115]{Left plot: sampled behavioural parameters (on log scale) for elk-115, $\lambda_1$ is the switching rate out of the `foraging' state and $\lambda_2$ is the switching rate out of the `travelling' state. Upper right plot: probability of residing in behaviour 2 (`travelling') over time. Lower right plot: probability of residing in behaviour 2 using the \texttt{R}~package \texttt{moveHMM}~\citep{MoveHMM}. In both plots on the right, points are included to highlight the times/frequency of observations.}
\label{Fig:ElkBehav}
\end{figure}

Samples from the posterior distributions for the two rates of switching defining the behavioural process are shown in the left panel of Figure~\ref{Fig:ElkBehav}. Posterior summary statistics for the switching rates are given in Table~\ref{Tab:ParamStat}, with the $90\%$ credible intervals leading to a mean residence time in state~1 being between 4--11~days and in state~2 between 10--36~hours. The behavioural parameters pass standard convergence diagnostics, with effective sample size of over 125. The right panel of Figure~\ref{Fig:ElkBehav} displays the probability of being in behavioural state~2 throughout the course of the sampling period. Additionally, the corresponding state probabilities estimated by fitting a hidden Markov model as in the vignette of \texttt{moveHMM} (but using the larger dataset of tracks from four elk) are shown below. The two models can be seen to identify the same areas of the movement path as being in the `travelling' state, however the residence times in this state differ between the two models, with the hidden Markov model classifying three long stays in state~2 in the middle of the observation period. 

\section{Discussion}
\label{Section:Discussion}
We have provided methodology for Bayesian inference for continuous-time, multistate movement. The behavioural process leads to a flexible range of movement patterns, whilst the continuous-time formulation allows missing and irregular observations to be handled with ease. Movement within a behaviour has some similarities with the velocity-based continuous-time model of \citet{Johnson2008} but is more intuitive, enabling a separation of speed and direction that matches empirical observations well. Parameter interpretation is simpler when separated in this way, describing aspects of movement such as a mean travelling speed and a volatility to the direction of movement. Although continuous-time models based on $(x,y)$ locations \citep{Johnson2008,Blackwell2015} could be applied, with post-processing to determine the distribution of speed and bearing, the covariance structure of such distributions, and hence the implicit shapes of the paths, will not be the same as that presented here. Ecological justification for such a covariance structure may be difficult or lacking, whereas our model is directly defined by these quantities and therefore initially motivated by ecological ideas.  

For a given state and time interval, the distribution of the change in direction given by our model will always be a wrapped Gaussian centred at zero. A von Mises distribution (often used in discrete models; \citealp{McClintock2012}) centred at zero is very similar to this, but a von Mises (or other circular) distribution centred at $\pm \pi$ is not. In fact, no natural continuous-time process for change in direction would lead to such a distribution when observed at regular intervals. Such a distribution would require the expected rate of change of bearing to be non-zero, leading to paths that consistently form loops. While this may be appropriate occasionally~\citep{Boakes2011} we do not feel it is realistic in our example or in most published applications. It seems more likely that such a distribution emerges only as an artefact of some other process e.g. ignored measurement error~\citep{Hurford2009} or attraction to a particular location. The classification of a foraging state with a mean turning angle of $\pm \pi$ in many discrete-time applications is therefore questionable. The ecological interpretation of a `foraging' state would be better modelled as having a uniform turning angle, such as $\sigma_\theta^2 \rightarrow \infty$ in our model. 

Modelling in continuous time allows us to consider movement/behaviour between observation times, something not possible in discrete time. The estimated residency rate of the travelling state in the elk example suggests that there are parts of the movement path where short sojourns of fast movement occur. In fact, $72\%$ of the sampled values from the posterior distribution of $\lambda_2$ lead to a mean residence time of less than the 24~hour sampling scheme. In Figure~\ref{Fig:ElkPaths}, it can be seen in a number of places that the reconstruction involves a switch in to and back out of state~1 between two consecutive observations. The exact time when these short (between observation) switches in behaviour occur vary over the sampled reconstructions, but their presence has high probability. There is therefore information in the observed locations indicating a behavioural sojourn has occurred, but 
the precise time of its occurrence
is very uncertain. Being able to extract such qualitative information on short term behavioural switches from observations, albeit with uncertainty, gives extra insight into the movement that is not possible when switches can only occur at the observation time scale. 

Although the approach for inference here is an approximation to the underlying continuous-time model, advantages remain over discrete time: behavioural switching can occur continuously in contrast to strictly at observation times and the parameters of the model are scalable (representing parameters of a continuous-time model) rather than `per observation time'. Reducing the refined time scale will provide a `better' approximation to the underlying model, but does come with a computational cost. Simulation experiments on the effect of varying $\delta t$ (details omitted here for brevity) show that great improvements to parameter estimation can be made against using only observations by augmenting as little as four locations between observation pairs. Improving the approximation with further refinement was found to increase accuracy of parameter estimation further, but incurred additional computation time. 

The methods described here assume that observation error is negligible. Extending this to observation error is easily implemented, included in the single behavioural method of \citet{Parton2016}. This simple model assumed normally distributed errors, independent in space and time. There is therefore a single additional parameter describing the observation error (a mean error of zero is assumed). An extension to the inference method described here allows for such a parameter to be sampled as a Gibbs step, and the path reconstruction method can be extended to include error around observed locations. Extending further to allow for errors to be correlated in time could also be implemented without difficulty.  

The augmentation approach furthers our aim for comprehensible inference. The ability to view examples of path reconstructions, such as in Figure~\ref{Fig:ElkPaths}, aids in understanding the movement type associated with a given combination of parameters. Sampling a large number of reconstructions displays the uncertainty in the times at which behavioural switches occur and can easily be used to estimate the space/resource use of the animal at the local scale. With the resolution of environmental covariates increasing, this information can be correctly combined with local scale movement rather than assuming that only the covariate values corresponding to directly observed locations are important. For discussion of the wider issues of linking movement and resource use, see for example \citet{Johnson2008b}.

We have assumed here that transition rates between behaviours are constant. It would be desirable to allow these to depend on spatial covariates~\citep{Morales2004} or on location itself. Depending on the duration of study, it may also be useful to allow varying rates with time, perhaps periodically to reflect daily or annual cycles. Both these extensions could be addressed, without any additional approximation, using the framework in \citet{Blackwell2015}, applied there to movement models directly based on location (rather than velocity or steps and turns) with heterogeneity in both space and time.
More generally, we could capture some more of the complexity of behaviour by including an additional `resting' state, likely to occur at particular times of the day, with low or zero speed and perhaps a high volatility to represent the `forgetting' of bearing while resting. We do not explore that approach further here, preferring to illustrate the key ideas as simply as possible. 

\noindent 
\textbf{Acknowledgements} The authors thank Th\'eo Michelot and anonymous referees for their helpful and insightful comments that have improved this text. 

\bibliographystyle{plainnat}
\bibliography{bibliography}

\appendix
\section{Technical details of the inference algorithm}
\label{AppSection:InferenceDetails}

\subsection{Conditional distribution for behavioural parameters}
\label{AppSubsec:BehavParam}
The full conditional distribution, $\mathcal{L}(\B{\Phi}_B \ ; \ \B{B},\B{\theta},\B{\nu},\B{z},\B{\Phi}_M)$, simplified as $\mathcal{L}(\B{\Phi}_B \ ; \ \B{B})$, is the posterior for a fully observed continuous-time Markov chain. Sufficient statistics for such a process are given by $a_i$, the total time spent in state $i$, and $b_{i,j}$, the number of transitions from state $i$ to state $j$. Given independent, exchangeable prior distributions of $\lambda_i \sim \text{Gamma}(c_i,\ d_i)$ and $q_{i,1},\ldots,q_{i,n} \sim \text{Dirichlet}(\B{f}_i)$, the posterior distribution is
\begin{align*} \label{Eq:BehavPosterior}
\lambda_i \ | \ \B{B} &\sim \text{Gamma}(c_i + \sum_{j=1}^n b_{i,j},\ d_i + a_i), \\
q_{i,1},\ldots,q_{i,n} \ | \ \B{B} &\sim \text{Dirichlet}(\B{f}_i + \B{b}_i), \quad \text{where} \ \B{b}_i=\lbrace b_{i,1},\ldots,b_{i,n} \rbrace.
\end{align*}

\subsection{Conditional distribution for movement parameters}
\label{AppSubsec:MoveParam}
The full conditional distribution, $\mathcal{L} \left( \B{\Phi}_M \ ; \ \B{\Phi}_B, \B{B}, \B{\theta},\B{\nu}, \B{Z} \right)$, simplified to $\mathcal{L} \left( \B{\Phi}_M \ ; \ \B{\Phi}_B, \B{B}, \B{\theta},\B{\nu} \right)$ when there is no observation error present, is given as
\[
\mathcal{L} \left( \B{\Phi}_M \ ; \ \B{\Phi}_B, \B{B}, \B{\theta},\B{\nu} \right)
	\propto
	\mathcal{L} \left( \B{\Phi}_M \ ; \ \B{\Phi}_B, \B{B} \right)
	\mathcal{L} \left( \B{\theta}, \B{\nu} \ ; \ \B{\Phi}, \B{B} \right),
\]
up to a constant. Above, $\mathcal{L} \left( \B{\Phi}_M \ ; \ \B{\Phi}_B, \B{B} \right) = \mathcal{L} \left( \B{\Phi}_M \right)$ is the density of the prior of the movement parameters and 
\begin{align*} 
	\mathcal{L} \left( \B{\theta}, \B{\nu} \ ; \ \B{\Phi}, \B{B} \right)&=
	\pi_\theta \left( \theta_1 \right)
	\pi_\nu \left( \nu_1 \ | \ \B{\Phi} \right)
	\prod_{i=2}^M 
	\pi_\theta \left( \theta_i \ | \ \theta_{i-1}, \B{\Phi}, \B{B} \right)
	\pi_\nu \left( \nu_i \ | \ \nu_{i-1}, \B{\Phi}, \B{B} \right),
	\\
	&\text{where} \ \theta_1\sim\text{U}(-\pi,\pi),
	\\ 
	&\pi_\nu \left( \nu_1 \ | \ \B{\Phi}_M, \B{\Phi}_B \right)
=  \sum_{i=1}^n \pi_\nu \left( \nu_1 \ | \ \B{\Phi}_M, s_0=i \right) \mathcal{L}\left(s_0=i \ ; \ \B{\Phi}_B \right),
\end{align*}
using the equilibrium distribution of the Ornstein-Uhlenbeck process for the initial speed likelihood. The conditional likelihoods in the product above are given by equations~\ref{Eq:SimBearingSpeed}.

\subsection{Conditional distribution for the unobserved refined path}
\label{AppSubsec:Path}
The full conditional distribution of a section of the refined path, up to a constant, needed for the Metropolis-Hastings step can be written as
\begin{align*} 
&\mathcal{L} \left\lbrace \B{B}^*,\B{\theta}^*,\B{\nu}^* \ ; \ \B{\Phi}, \mathcal{F}, \B{Z}(b) \right\rbrace \notag \\ 
&= \mathcal{L} \left\lbrace \B{B}^*, \B{\theta}^* \ ; \ \B{\Phi}, \mathcal{F}, \B{Z}(b) \right\rbrace 
		 \mathcal{L} \left\lbrace \B{\nu}^* \ ; \ \B{B}^*, \B{\theta}^*, \B{\Phi}, \mathcal{F}, \B{Z}(b)\right\rbrace \\
&\propto \mathcal{L} \left\lbrace \B{B}^*, \B{\theta}^* \ ; \ \B{\Phi}, \mathcal{F} \right\rbrace
		 \mathcal{L} \left\lbrace \B{Z}(b) \ ; \ \B{B}^*, \B{\theta}^*, \B{\Phi}, \mathcal{F} \right\rbrace
         \mathcal{L} \left\lbrace \B{\nu}^*  \ ; \  \B{B}^*, \B{\theta}^*, \B{\Phi}, \mathcal{F}, \B{Z}(b)\right\rbrace 
\end{align*}
where $\mathcal{F} = \lbrace B(a), B(b), \theta_0, \theta_n, \nu_0, \nu_n, \B{Z}(a) \rbrace$. The simulation method employed to create a proposal for a refined path section, described in Section~\ref{Subsection:PathPropMethod}, results in a proposal distribution proportional to $\mathcal{L} \left\lbrace \B{B}^*, \B{\theta}^* \ ; \ \B{\Phi}, \mathcal{F} \right\rbrace
		 \mathcal{L} \left\lbrace \B{\nu}^* \ ; \ \B{B}^*, \B{\theta}^*, \B{\Phi}, \mathcal{F}, \B{Z}(b)\right\rbrace$, and so the Metropolis-Hastings acceptance ratio is based only on the marginal distribution $\mathcal{L} \left\lbrace \B{Z}(b) \ ; \ \B{B}^*, \B{\theta}^*, \B{\Phi}, \mathcal{F} \right\rbrace$ given in Equation~\ref{Eq:JointStepLoc}.

\end{document}